\documentclass[12pt]{article}

\usepackage{jinstpub}

\usepackage{hyperref}

\title{\boldmath Amplitude Walk in Fast Timing: The Role of Dual Thresholds}

\author[a]{Sebastian~White}
\author[b]{Alessio~Boletti}

\emailAdd{sebastian.white@cern.ch}

\affiliation[a,b]{  Laboratory of Instrumentation and Experimental Particle Physics (LIP), Portugal}

\abstract{ 
We apply lessons from fast timing detector R$\&$D to strategies for initial calibration of large timing arrays
at future colliders. Detector R$\&$D often benefits from detailed information about the sensor and front-end
signal (waveform capture) as well as a quality time reference and tracking.
On the other hand, the systems for charged particle (MIP) timing under construction for the CERN High Luminosity
LHC log only limited information for each timing channel- usually amplitude and the time of the leading edge.
Furthermore the high event rates certainly present a challenge for \textit{in situ }calibration of the large (compared to intrinsic)
time jitter of the leading edge with pulse amplitude- amplitude walk.
In the examples presented here we find a simple linear dependence of walk on the inverse of the pulse slope at
threshold for the dynamic range (in amplitude) suitable to charged particle timing. We present a straightforward 
calibration method for the small variation in the corresponding
coefficient from channel-to-channel.}
	
\keywords{ Timing detectors, SiPM, Signal Processing}

\begin{document}
\maketitle
\flushbottom
\section{Introduction}
\label{sec:Introduction}

	Recent advances in timing sensor technology have enabled the construction of large ($\sim10^5$ channel) timing detector arrays for the High Luminosity upgrade of CERN's
Large Hadron Collider (LHC). We focus here on those designed to measure charged particle (i.e. MIP) time of arrival, while at least two calorimeters aim to add timing capability to their energy measurement.

	The MIP timing detectors \cite{TDR} have demonstrated an initial $\sim30$ picosecond time resolution degrading by roughly a factor of two due to radiation damage over their
useful lifetime. This remarkable \textit{intrinsic} performance brings to the fore the significant calibration effort required to exploit this resolution- particularly on Day-1, when it is
at its best.
 
 	Since the beginning of the LHC, the ALICE experiment has incorporated a time-of-flight (TOF) system with $\sim150$k channels and achieved a charged particle time resolution of 82 picoseconds initially, and starting from LHC Run-2 improving to 52 picoseconds \cite{ALICE2} \cite{ALICE}, largely due to improvements in the walk correction.
	
	The calibration effort, in the ALICE case, was not insignificant and many aspects directly bear on the likely procedures to be employed in the new systems of ATLAS and CMS
upgrades. Nevertheless, the ALICE calibration procedure benefitted from the fact that, by comparison, ALICE is a low rate experiment. The ALICE TPC tracking system requires an interaction rate so low that events only capture a single interaction whereas the ATLAS and CMS experiments are designed to handle a 140-200 p-p interaction ``pileup" \cite{Shiltsev}. This alone
is a strong incentive to factor the three interdependent  calibration activities which resulted in ALICE's performance.

	The ALICE activities consist of  \cite{ALICE2}:
\begin{itemize}
\item Establishing  a ``time of event" with three different tools. 
\begin{enumerate}
\item  In an ideal world the colliding beam interactions would be restricted to a point in space and time by the collider RF. In fact, the interactions will be spread in time by the finite bunch length (to $\sim180$ picoseconds rms) and correspondingly in space. The collider clock signal,  whose arrival at ALICE is subject to correctable thermal drifts, is useful to this level of accuracy.
\item For a fraction of events (lower in p-p collisions) forward t$_0$ counters capture the ``time of event".
\item When there are more than 3 tracks fully reconstructed in the TOF detectors these hits are propagated back to a vertex using track information and the assumption of $\pi^\pm$ identity.
\end{enumerate}
\item Starting from the collider RF reference, channel-to-channel offsets due to propagation delays are measured and tracked over the life of the experiment.
\item In ALICE, as in the MIP timing detectors for the LHC upgrades, the TOF system records leading edge times as well as ``time-over-threshold"-TOT (which is proportional to the pulse amplitude).
Therefore, in parallel with the above calibrations, an amplitude walk correction- i.e. a functional dependence of time walk versus amplitude is developed for each channel. In the ALICE case this corresponds to finding the terms in a fifth (and later higher) order polynomial for each channel.
\end{itemize}
		
			Considering the ATLAS and CMS HL-LHC upgrades, the high pileup environment will certainly complicate the above procedure. The ``time of event" reference will
certainly consist of a more diffuse collection of 1-2$\times10^2$ separate event times (which are still referred to as ``beam spot" in calibration discussions).

	In any case, there is good reason to explore a strategy for developing an amplitude walk correction for all timing channels which is stand-alone. This would decouple the correction from event time determination and charged particle tracking tools. It would focus on the direct dependence of leading edge time walk vs. amplitude on the properties of the timing channel itself. 
	
	Though
it is not directly relevant for the systems we are discussing (which employ leading edge timing) there have, in the past, been approaches to timing electronics that aim to remove walk
directly \cite{Knoll} \cite{Leo} with names that are mostly self-explanatory (i.e. zero-crossing, Constant Fraction and Amplitude and Risetime Compensated timing).
Instead the field has moved to an approach where both amplitude and leading edge time are captured, which will be justified if software correction ultimately leads to higher resolution. 	
	
	An earlier paper \cite{linear} discussed laboratory data where the Silicon PhotoMultipliers (SiPMs), illuminated by a pulsed laser, were read out using a linear front end producing full waveforms recorded by a digital oscilloscope. For this case the linearity of the system enables scaling directly to an arbitrary waveform amplitude. It was shown that ``ideal constant fraction timing" could be achieved so long as a single calibration at the selected amplitude (relating amplitude to slope at threshold) was available. 
	
	This is stand-alone in the sense that, once the single calibration is achieved, an algorithm using only leading edge time and amplitude in the data can be shown to coincide exactly with the walk that you observe when using the laser `time of event" information. An important characteristic of those data was that the system provided a linear response in the sense that any waveforms, when re-scaled by amplitude, would lie on top of one another. 
	
	For the rest of this paper we will focus on the Barrel Timing Layer (BTL) component of the CMS MIP timing detector upgrade \cite{TDR}. The BTL detector, as further described below, consists of arrays of LYSO:Ce scintillator bars coupled to SiPMs at both ends of the bars.
We focus on lab data taken with the full BTL system. For these data, the LYSO crystals were excited by a pulsed UV laser reproducing the signals corresponding to MIP response and the SiPM signals are processed by the final version of the BTL readout electronics, referred to hereafter as TOFHIR (the final production model used for this paper is ``TOFHIR2C"). 
	
	A feature of the TOFHIR ASIC is that more than
one leading edge time can be captured per event. This enables periodic calibration data runs wherein tables of pulse slope at threshold vs amplitude can be compiled. The calibration run
and the correction procedure that uses the calibration are both stand-alone in the above sense. We find, for a set of 25 different channels a simple linear dependence of the walk correction on the inverse of the pulse slope at threshold. The coefficients of this linear dependence can also be captured from the calibration data.

	In order to evaluate the calibration procedure that we propose, we can directly compare to the more standard procedure (exemplified by ALICE), where a successful walk calibration implies that reference to a precise event time shows that variation with amplitude has been eliminated. This will be done in Figures 2 and 9. It should be noted that overall time resolution
is a less direct evaluation since intrinsic time resolution generally also depends on amplitude (due to photostatistics, signal-to-noise ratio, etc.).
\subsection{Outline}

	Motivated by simpler examples we demonstrate that, also for the ASIC under study, the amplitude walk calibration reduces to a simple linear expression relating inverse of the pulse slope at threshold to walk.
	
	A single coefficient gives the relation between ``corrected time" and the time and slope at threshold measured in a given channel.
Of course, reducing initial calibration to a single, possibly channel dependent, constant is in itself an advantage.
	
	From the sample of channels investigated below, we find that the calibration constants do vary among channels ($\pm 22\%$) but this variation can be taken into account
from the initial stand-alone calibration run (where slope and charge are recorded).

	The above plan to put in place ``Day-1" walk calibration never uses an external time reference (t$_0$). It assumes that the following study gives a complete prescription
for initializing the constants of every channel.

	In order to arrive at this prescription we, of course, employ a stable time reference in the laser setup to validate this prescription. But ``Day-1" calibration is based
only on data internal to the detector subsystem (the CMS Barrel Timing Layer (BTL) \cite{TDR} in the case discussed below).

	If we are successful in this program, the "pulse slope at threshold" will give the same quality calibration as an ideal measurement with known time reference (see demonstration in
Figure 9). This provides an enormous benefit since it obviates the need for the ideal clean environment needed for the usual brute force calibration.

\section{Slope Measurement with dual thresholds}
		What is the alternative? One particular ASIC developed for the BTL- the TOFHIR \cite{tahereh}- incorporates an additional timing discriminator whose threshold
setting can be adjusted to slightly differ from the first discriminator as in Figure 1 (left). The threshold settings are given by the corresponding DAC values which control them. If the second threshold
is set just above the first -see Figure 1 (left)- by $\delta$I (the TOFHIR uses current discriminators, so  ``thr"  implies  $I_{threshold}$ in this paper) then:

\begin{equation}
Slope= \frac{ thr_2-thr_1}{ dt}
\end{equation}

\begin{figure}[!htp]
\centering
\includegraphics[width=.5\textwidth]{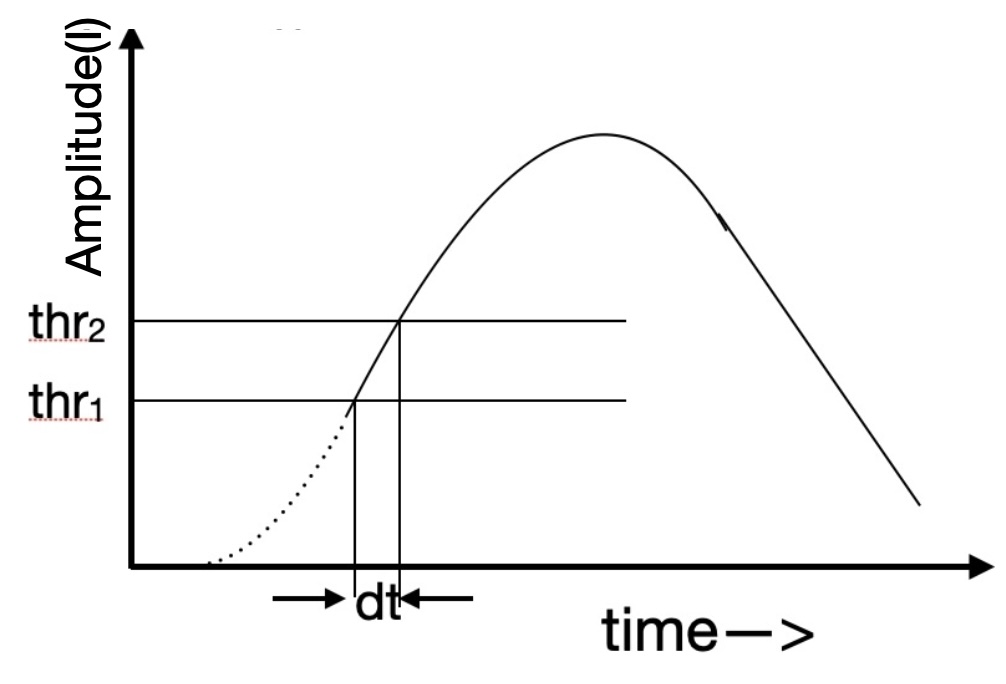}
\includegraphics[width=.45\textwidth]{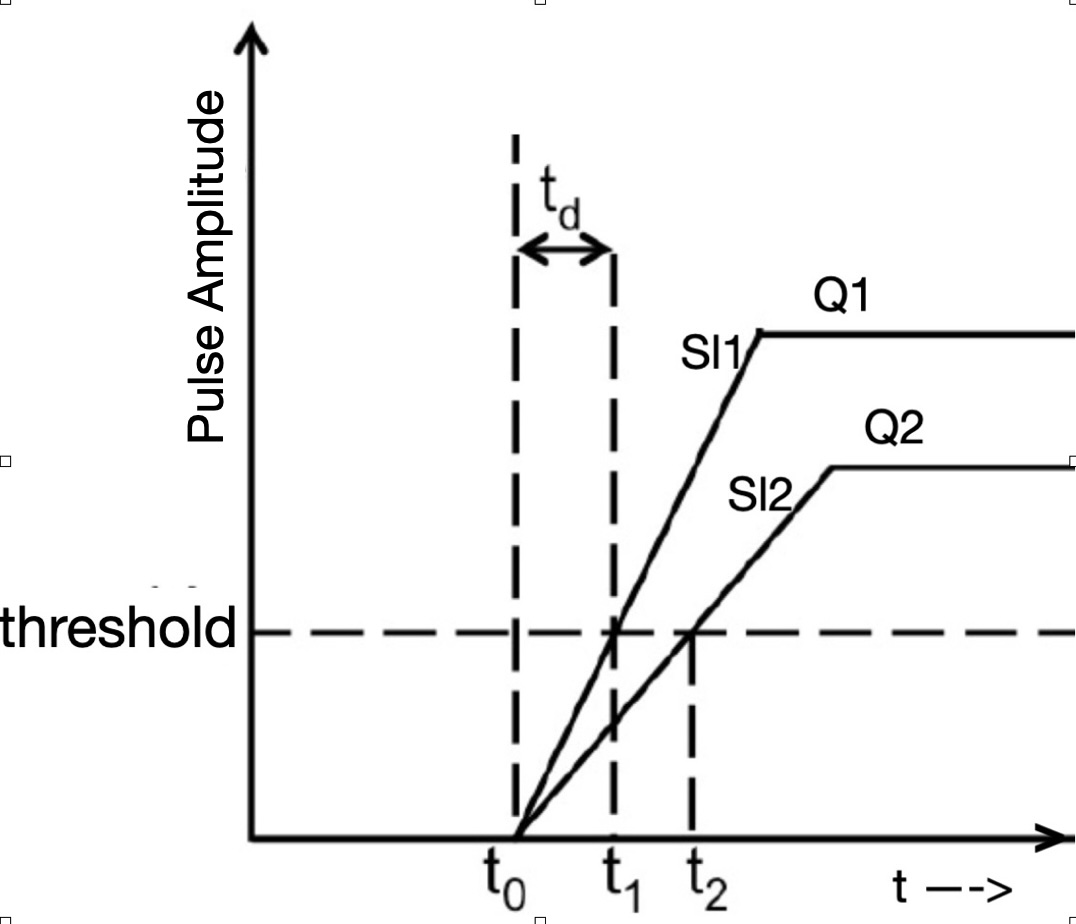}
\caption{ Obtaining  ``pulse slope at threshold" from dual thresholds as used throughout this paper (left). A simple example (right)  illustrating the direct relation between ``pulse slope at threshold" and walk for two different signals with slopes Sl1 and Sl2 and amplitudes Q1 and Q2.}
\label{fig:naive}       
\end{figure}

		The procedure we propose for walk correction does not require event-by-event recording of the slope. The walk correction deals with the average behavior of a channel- in
particular benefiting from the stability of the relation between amplitude and slope. One need only record data with sufficient statistics to capture the slope corresponding to a given
value of amplitude from an ensemble of slope measurements.

\subsection{Naive case}

	We wish to correct to the same time of arrival irrespective of amplitude. We illustrate the relation between walk and pulse slope at threshold with the simple example in Figure 1 (right). Consider two pulses given equal time of arrival but differing amplitudes (Q1,Q2).
In this example the slopes (Sl1, Sl2) are clearly proportional to the corresponding amplitude (i.e. charge, Q). We will see later that, more generally, slope and Q are usually linearly related but with a possible offset.

	The slopes (Figure 1, right) are given by:
		
\begin{equation}
Slope_i= \frac{I_{th}}{t_i-t_0}
\end{equation}

so the general rule for correcting to a common time, independent of Q is:

\begin{equation}
t_0= t_i-\frac{I_{th}}{Slope_i}
\end{equation}

	Since, in the remainder of this paper we will discover that the correction for walk is generally a linear expression in inverse slope we will adopt the notation Amplitude Walk Coefficient (AWC- with the same dimensions as threshold) and re-write eqn. 2.3 as:
	
\begin{equation}
t_{corrected}= t_i-AWC*\frac{1}{Slope_i}
\end{equation}
\subsubsection{Lessons from Naive Case}

	
The naive case illustrates the fact that amplitude walk is all about pulse slope\footnote{We avoid the term ``slew rate" in this context since the usual use in amplifier characteristics has a different meaning.} and the calibration may reduce to the determination of a single parameter-the AWC.
The common element in timing ASICs -i.e. recording pulse amplitude or time over threshold - should really be viewed as an indirect measure of the slope. We will see that the dual
threshold feature, from which we can measure slope, simplifies the calibration.

\subsection{Departure from the Naive Case}

	The key criterion for applicability of the method we propose is linearity.
\begin{itemize}
\item{In the simplest case- the naive example above (Figure 1, right)- the timing pulse leading edge is simply a straight line. In practical circuits, bandwidth limits the transition from baseline
but this has a minor effect on this example. So in this case one needs only knowledge of the slope for a given event.}
\item{More generally, linearity refers to the system response. If response is linear then waveforms of different amplitudes will be identical so long as they are re-scaled by amplitude. For this case,
the benefit of recording the pulse slope was demonstrated in an earlier report \cite{linear} and the effectiveness is illustrated in Figure 2, which is taken from that report. So again, for this case one only needs knowledge of the slope for a given event.}
\item{ Lastly, and for the remainder of this paper, we deal with a case where neither notion of linearity applies- the TOFHIR ASIC\footnote{an example of non-scaling is evident from Figure 4.}. We will see that, in this case, pulse slope and amplitude
information complement one another, yielding once again a single parameter correction- the AWC.}
\end{itemize}
	
\begin{figure}[!htp]
\centering
\includegraphics[width=.45\textwidth]{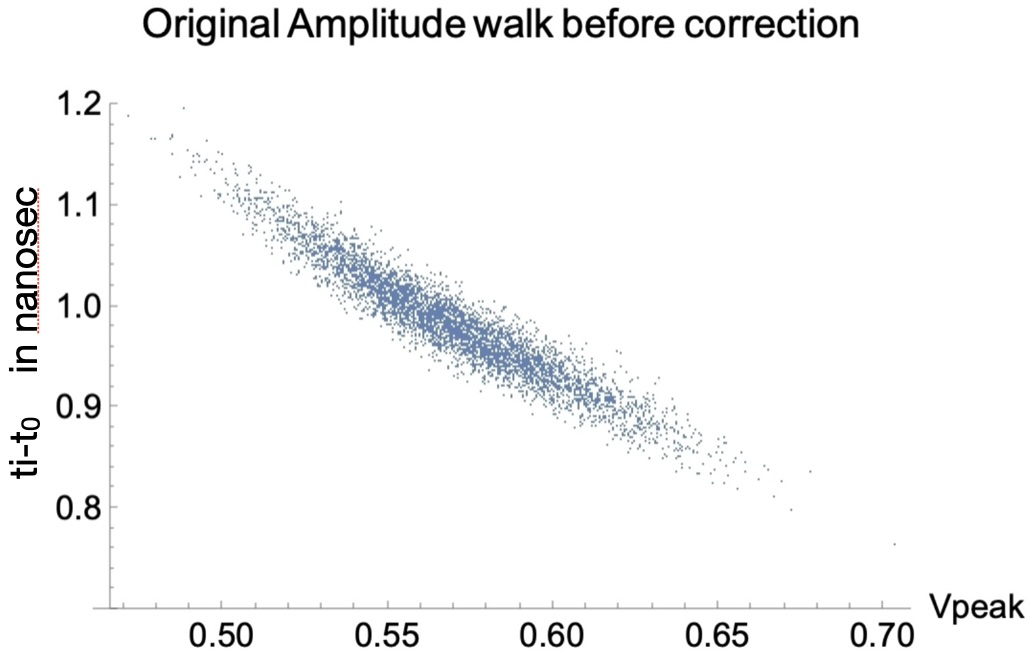}
\includegraphics[width=.45\textwidth]{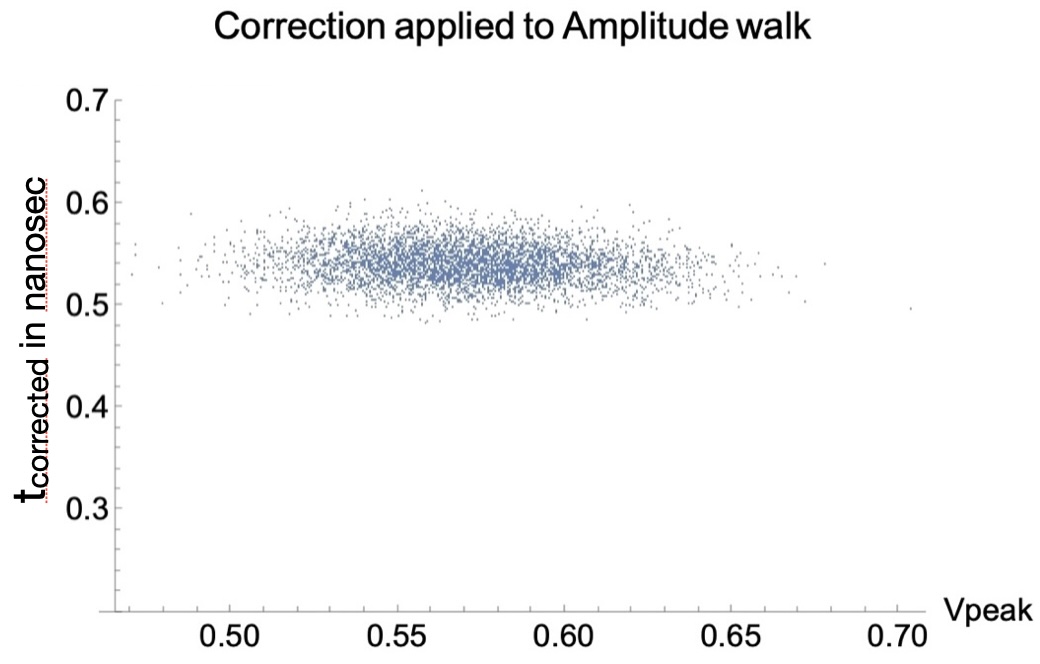}
\caption{The resulting walk corrected data (right) after application of an analytic expression in pulse slope for a system with linear response- taken from ref. 7.}
\label{fig:walk}       
\end{figure}

\section{Data Sets Used for this Paper}
\label{sec:Data}

	The laboratory data on which the following analysis is based were recorded using a test stand at the PETSYS company which produced the TOFHIR ASIC under contract with CMS.
The basic setup, shown in Figure 3, consists of a fast pulsed UV laser\footnote{NKT Photonics PIL1-037-40 UV laser.} that sequentially illuminates an array of 16 LYSO bars. The LYSO
array is a prototype for the CMS BTL MIP timing detector (MTD) upgrade described in ref. \cite{tdr2}. Each 54.7$\times 3.0\times 3.75mm^3$ bar has Hamamatsu  SiPMs\footnote{Hamamatsu  S15408 MPPC array for CMS BTL Type Custom} coupled to both ends which are read by a single 32-channel TOFHIR ASIC. Each LYSO bar is wrapped in reflective tape to prevent optical crosstalk except for a small opening that allows light from a 375 nm pulsed laser to directly excite the LYSO bar.

\begin{figure}[!htp]
\centering
\includegraphics[width=.45\textwidth]{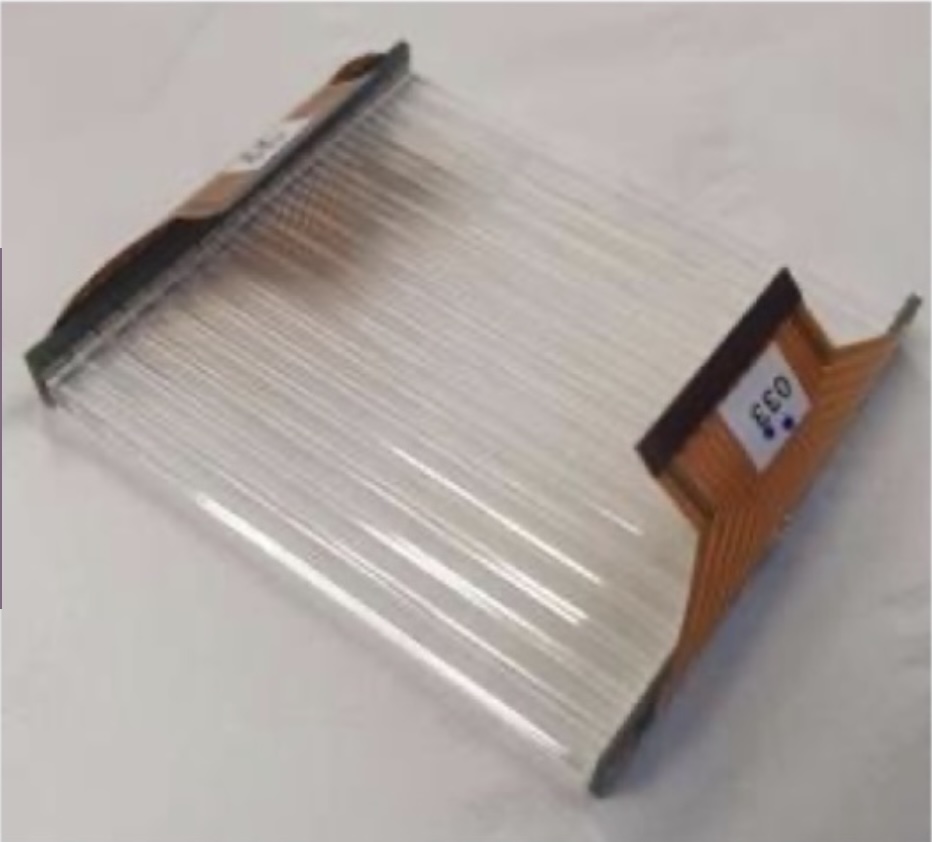}
\includegraphics[width=.5\textwidth]{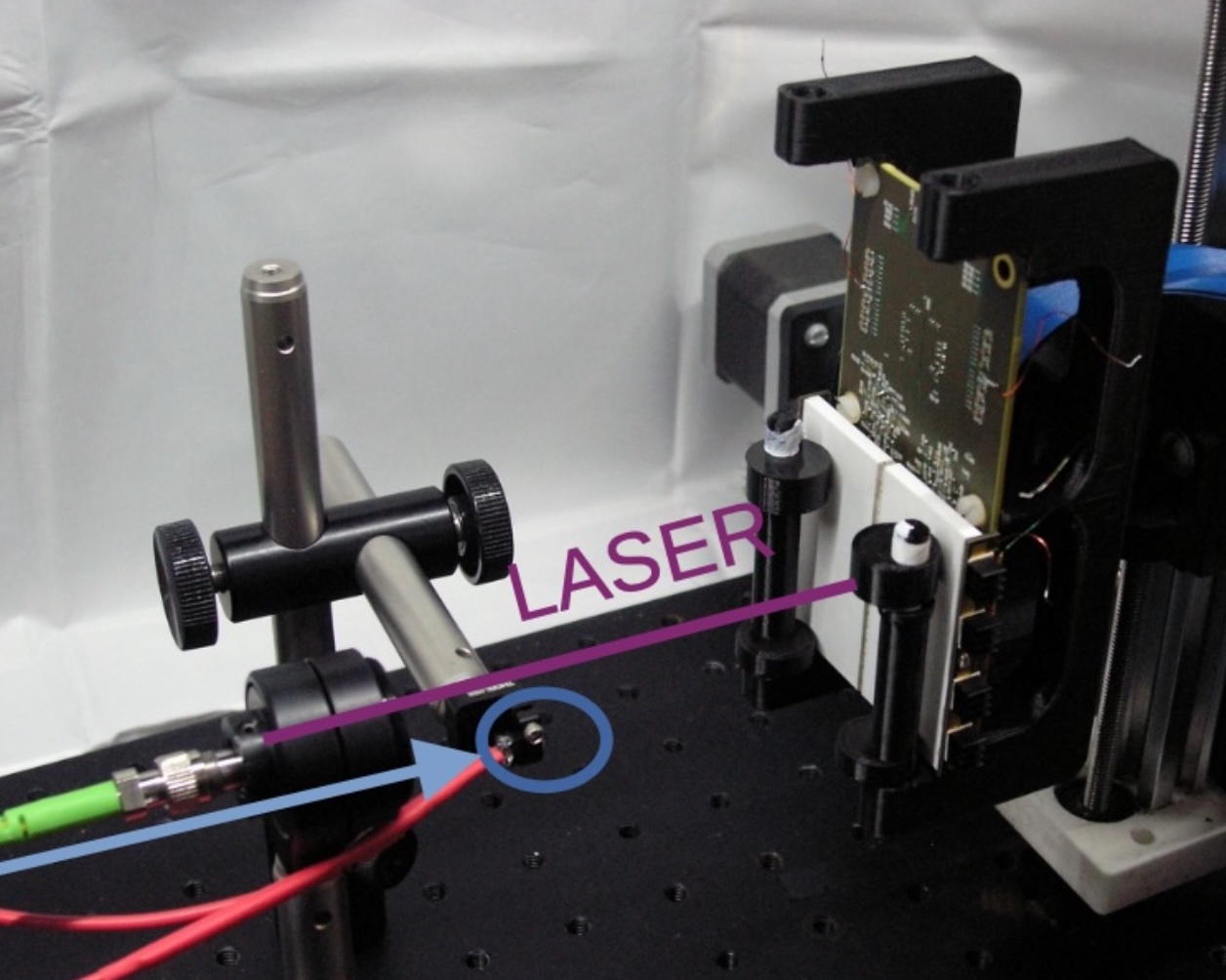}
\caption{A single LYSO/SiPM  array (Left). The light reflector wrapping has been removed in a single strip (Right) so that a pulsed UV laser beam can illuminate individual bars.}
\label{fig:setup1}       
\end{figure}

	Further details of the LYSO/SiPM properties can be found in ref. \cite{tdr2} but, for the purposes of this paper, it is sufficient to keep in mind that the detector is intrinsically linear in
response so the departure from overall linearity of the system is attributed to the ASIC itself.  The resolution of TOFHIR measurements differs significantly for amplitude (Q) and slope.
The amplitude measurement is the output of a charge ADC integrating a signal copy clipped to 12 nanoseconds and has a resolution of $3\%$ rms. The slope measurement derives from the two current discriminators (T$_1$ and T$_2$). The threshold settings are in themselves precisely set by DACs and periodic calibration of the baselines. However, as discussed in the Appendix, residual electronic noise limits the slope measurement in a given event to $\sim10-15\%$. Throughout this paper the Q to slope correspondence lookup tables are understood to capture $\geq100$ slope measurements for each Q bin, in order to get a stable mean value.

	The TOFHIR employs effectively single delay line shaping using R-C networks internal to the
chip and many aspects of its performance are captured in the CADENCE tool which was used in the design. However, given our limited access to the CADENCE tool used by PETSYS, it hasn't been possible to reproduce the non-linear aspects which complicate the following analysis.

	In this respect, we develop a data driven solution within which several features can be justified from simple considerations- for example, that the lookup tables generated by calibration runs capture useful indicators of trimable channel differences.

\subsection{Laser Data}

	In a particular dataset we step through all 16 bars, one at a time. For each bar we step through a sequence of  6 laser intensity settings which cover the SiPM output range that matches
the TOFHIR ADC limits when the SiPM bias is matched to the expected conditions at the start of operations-i.e. ``Day-1 conditions".

	Finally, the laser pulses are repeated several hundred times at each setting of the TOFHIR timing threshold (we are scanning timing discriminator 2- hence the labels ``thr2"), which spans from roughly the baseline to several tens of $\mu$Amps. The threshold setting is controlled by a DAC with 0.3125 $\mu$Amp least count and the baseline is measured separately for each channel.

	Following this procedure we obtain the data plotted in Figure 4 (left) for a given channel. There are 6 laser intensities in this figure and for each intensity the points accumulated from $\sim 200$ laser pulses at each threshold setting map out the leading edge in the neighborhood of the expected useful threshold (typically 4-6 $\mu$Amp).
	
	Referring to the 6 leading edge profiles in Figure 4 (left), it is easily verified that the system does not obey the linearity (i.e. scaling) behavior defined in Section 2.2 (i.e. the curves,
	when re-scaled by Q, do not coincide).

\begin{figure}[!htp]
\includegraphics[width=.45\textwidth]{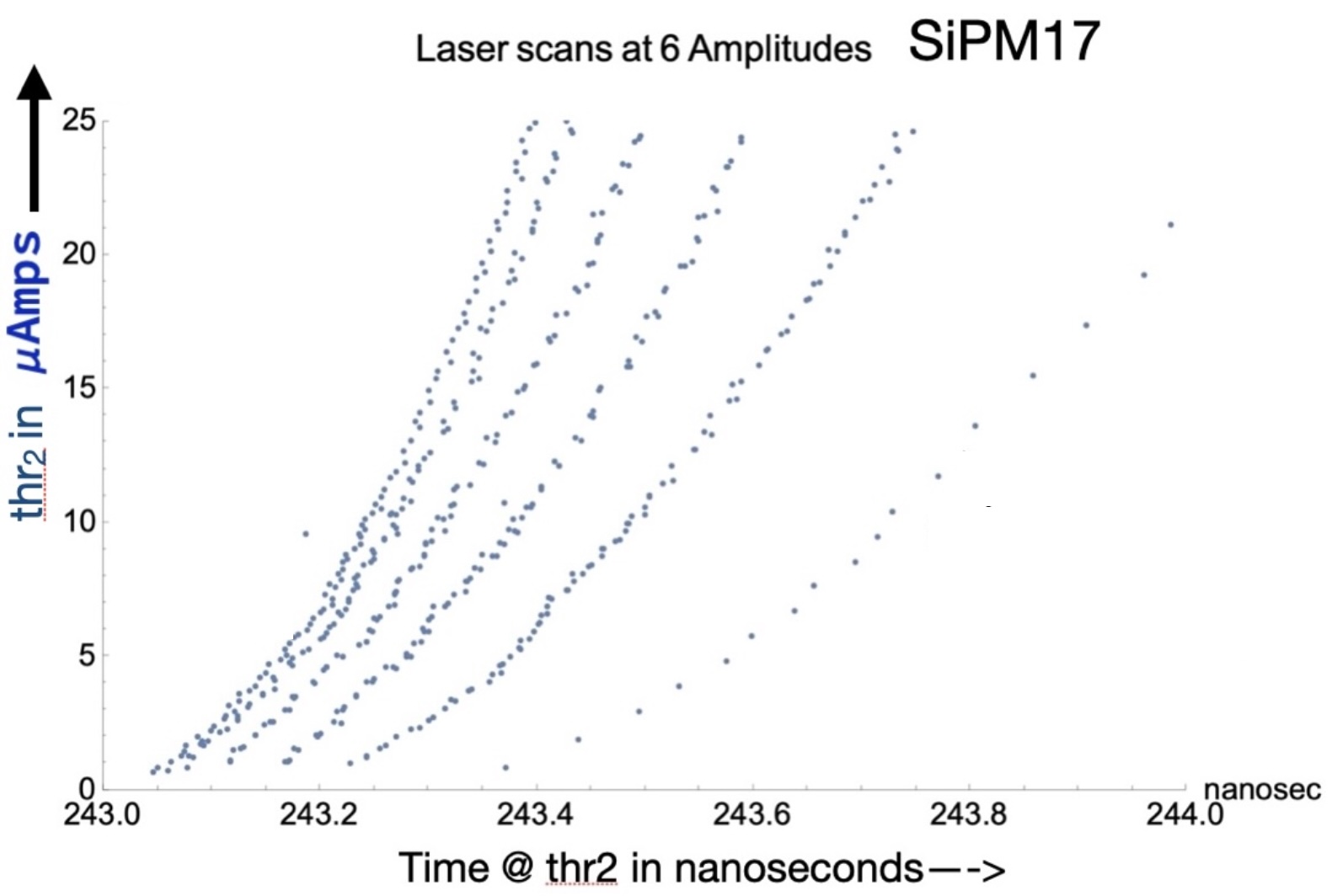}
\includegraphics[width=.5\textwidth]{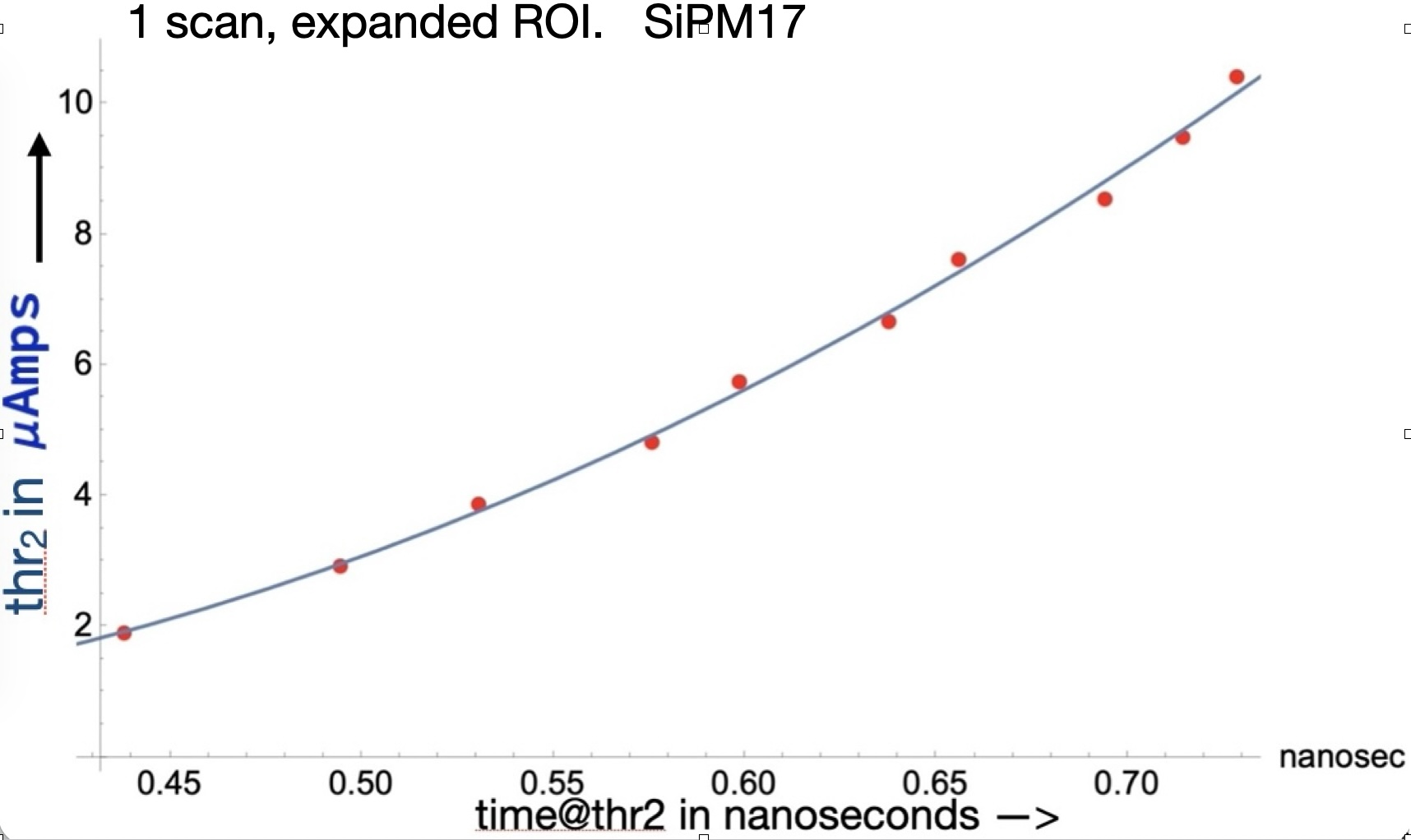}
\caption{A typical data set for one channel (SiPM17) , where each point is an average of $\sim200$ laser shots (left) and the resulting curves represent the pulse shape at the input
of the timing discriminator for each laser intensity. Since
we present results for a range of thresholds, we capture times (i.e. walk relative to t$_{corrected}$) and slopes in a quadratic fit (right) for this analysis.}
\label{fig:scan}       
\end{figure}

	We can use the scans in Figure 4 to capture the pulse slope and timestamp (i.e. $t_i$ of eqn 2.4) at threshold for the 6 pulse amplitudes. For the remaining discussion we take the measured points for each channel\footnote{of the 32 input channels, 2 were disconnected and 5 did not cover the full range of thresholds.} and perform quadratic fits (see Figure 4-right) in order to capture time and slope in a range of threshold settings. Also, in Figure 4-right, we remove the arbitrary $\sim 242$ nanosecond offset of the laser time stamp.

	For each channel we map out the relation between inverse slope and time at threshold at 6 different intensity settings. The results for a typical channel are displayed in Figure 5. In all cases the data are well described by a linear dependence of time at threshold ( $t_i$) on the inverse slope and we conclude that the TOFHIR data for a given threshold follow the simple form of eqn. 2.4. The slope of the fitted lines correspond to the coefficient ``AWC" we defined above.

\begin{figure}[!htp]
\centering
\includegraphics[width=.75\textwidth]{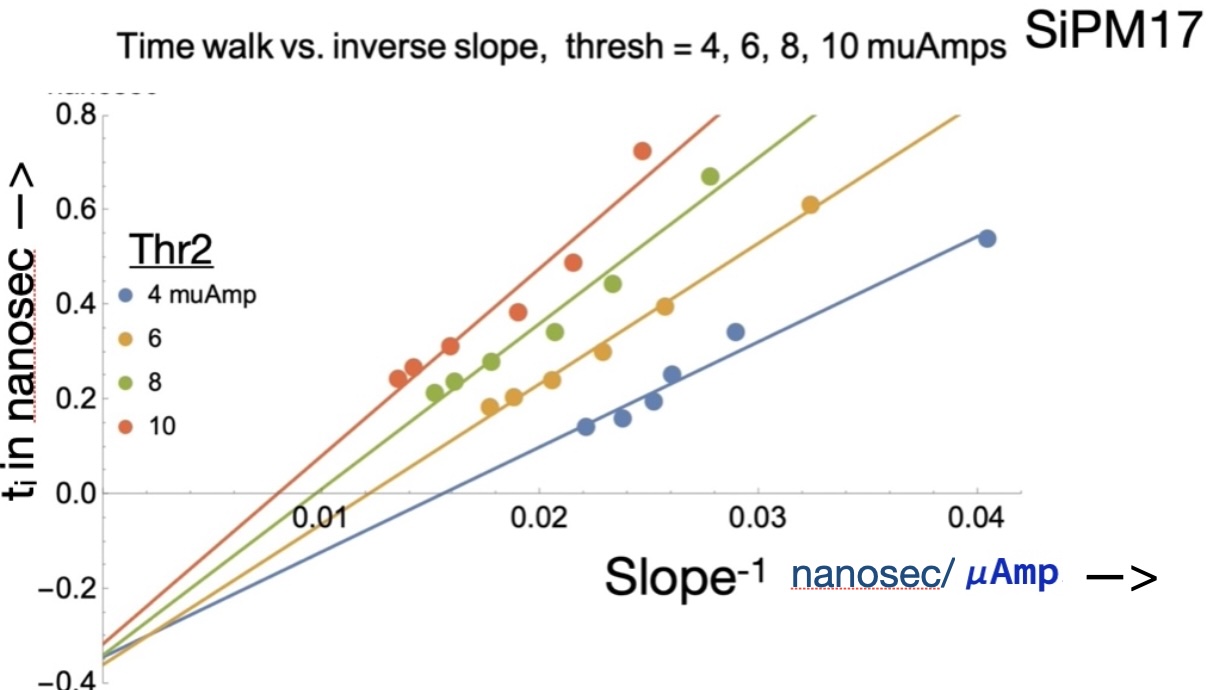}
\caption{The time at threshold ( i.e. $t_i$  in eqn 2.4) vs. inverse slope are well fitted to a linear relation in this and all other channels. Measurements are repeated for 4 different  ``thr2" settings.}
\label{fig:awc}       
\end{figure}

	Unlike the linear examples discussed in Section 2.2  ( first 2 bullets), we have no a priori expectation for the value of the Amplitude Walk Correction. So we next ask the question ``what is the spread in
AWC values" for our sample of 25 channels. More importantly, since we are looking for a procedure to calibrate a much larger sample, ``can we anticipate individual AWC values using TOFHIR-only calibration data". The answer appears to be ``yes".

\section{ Determining individual Amplitude Walk Coefficients}
\label{sec:AWC}

	Let us take 6 $\mu$Amps as a threshold setting and examine the fitted values for AWC among the 25 channels. The result is plotted in Figure 6. The spread in values is rms$\sim 22\%$, which
might be regarded as acceptable for an initial calibration. Nevertheless the crude lookup table of Slope vs. Q generated for the initial calibration reveals the origin
of spread in these coefficients and provides a method to compensate for this spread.

\begin{figure}[!htp]
\includegraphics[width=.45\textwidth]{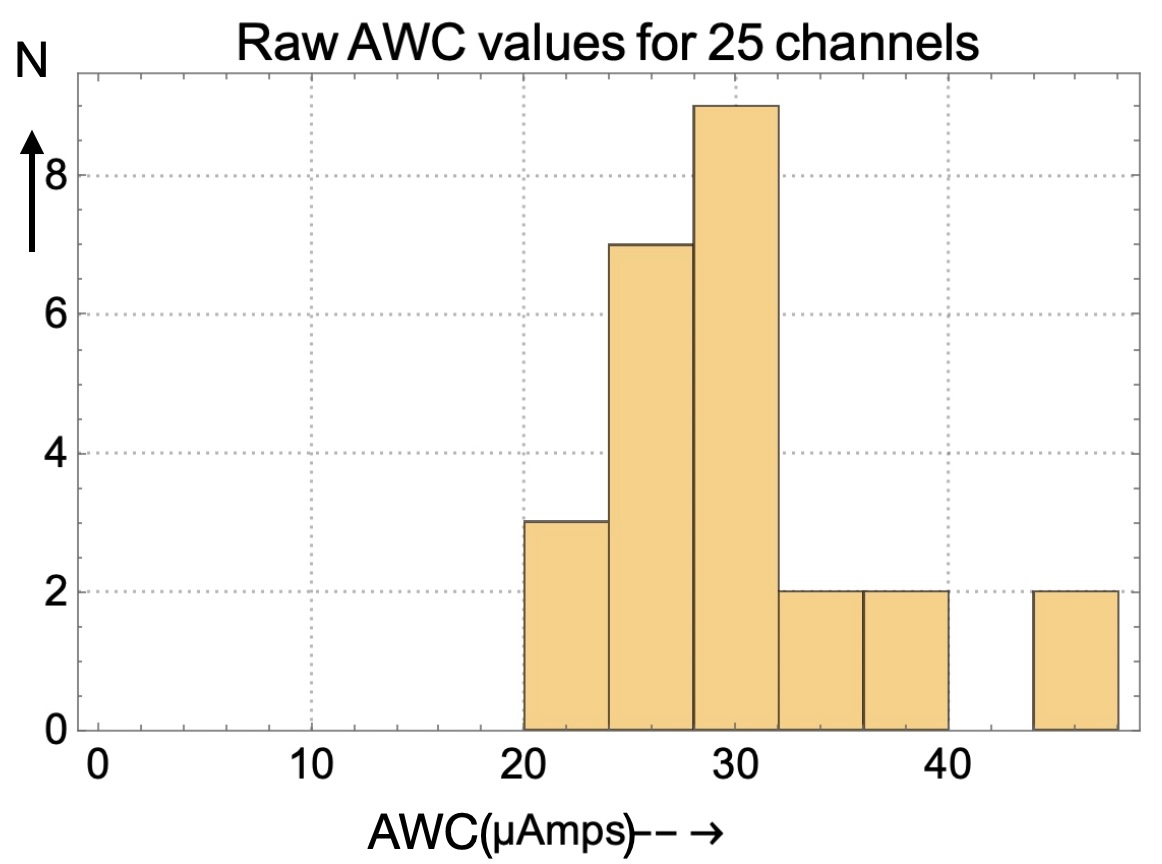}
\includegraphics[width=.45\textwidth]{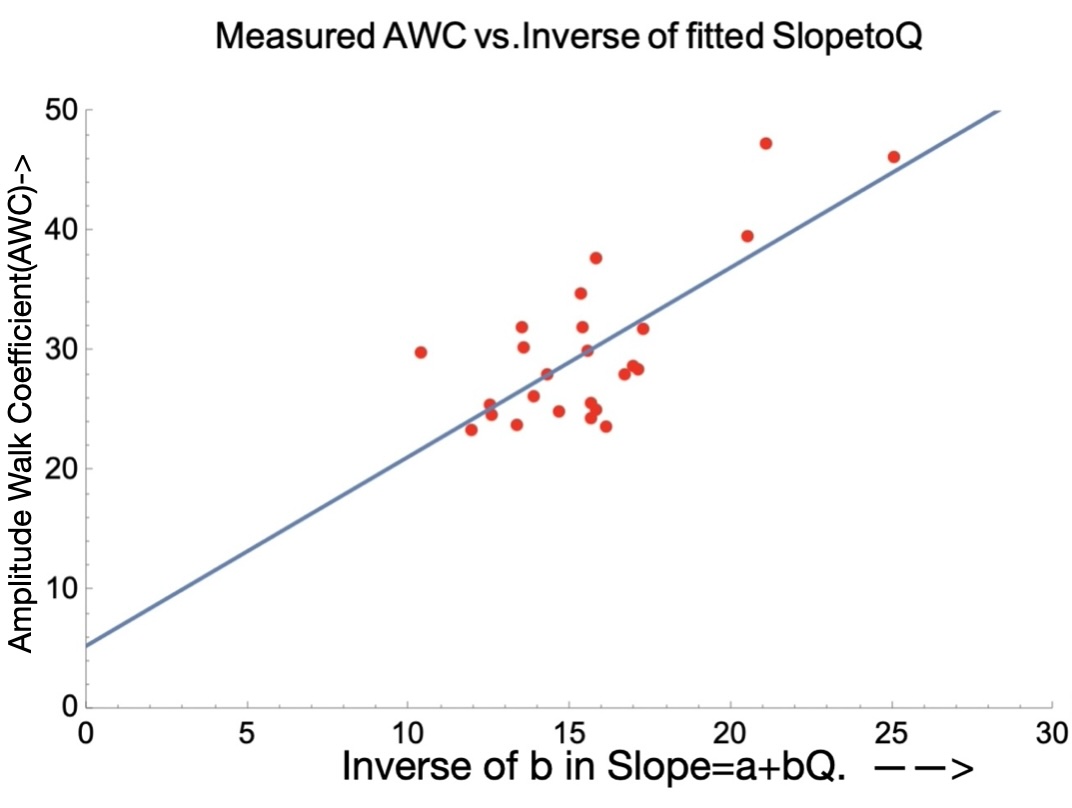}
\caption{ The slopes (left) fitted, as in Figure 5, at 6 $\mu$Amp threshold have an rms spread of $\sim22\%$ (left). However this spread is correlated with a variation in Slope-to-Q ratio 
captured in calibration data- i.e. the b parameter derived from Figure 7. Using this correlation we reduce the uncertainty on AWC to $\sim13\%$ (right).}
\label{fig:corr}       
\end{figure}

	Figure 7 is an example of the measurements from which the table is constructed for a particular channel and with a threshold setting of 6$\mu$Amps. The average value of the line in Figure 7 and its progression with threshold (8 thresholds from 3 to 10 $\mu$Amps are plotted in Figure 8) will be used in Eqn 4.1 as a measure of a particular channel's Mean Slope to Q Ratio ("MSlQ"). Where "MSlQ" is the value (b$_i$) obtained by fitting the data in Figure 7 to the form:
	
\begin{equation}
Slope_i=a_i+b_i\times Q
\end{equation}

\begin{figure}[!htp]
\centering
\includegraphics[width=.55\textwidth]{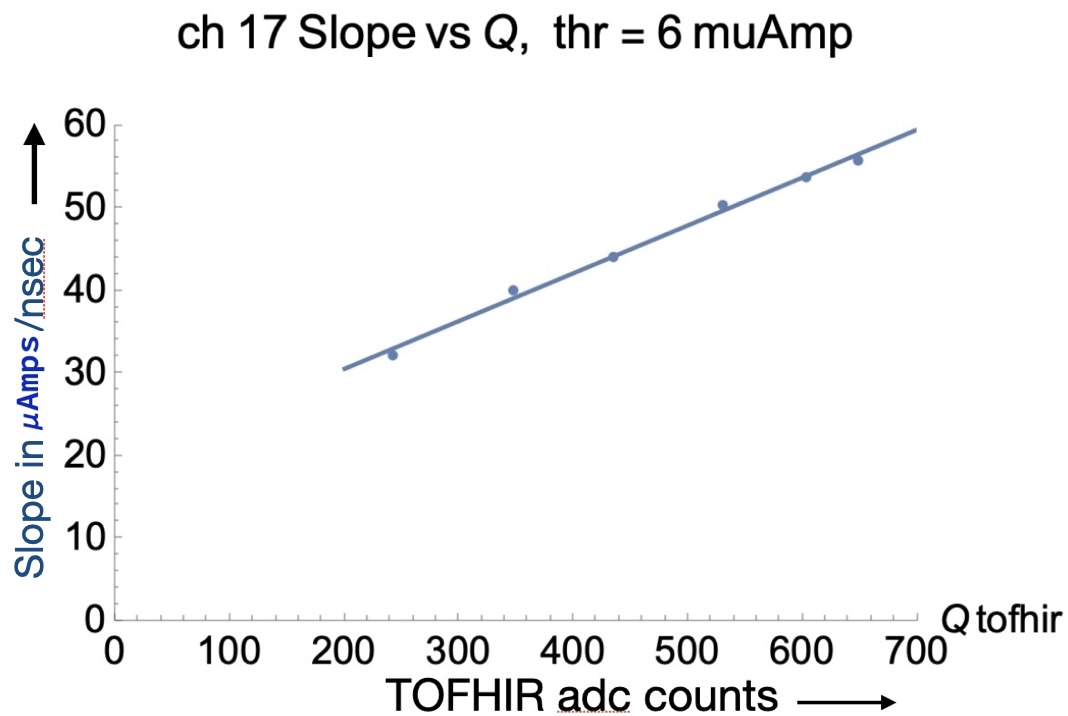}
\caption{The mapping from Q- the ADC output- to Slope for a given channel at 6 $\mu$Amp threshold.}
\label{fig:slQ}       
\end{figure}

	Assuming the calibration data are taken at several threshold settings we can determine the correlation between slope and Q for every channel, in particular how rapidly the slope develops with increasing Q. If this differs among channels then
it is reasonable to expect that channels with a more rapid evolution of slope would have a smaller coefficient. This turns out to be the case as can be seen from Figure 8, where channels
with smaller values of AWC clearly have a more rapid development of Slope vs. Q. In the following we will use simply the "MSlQ" fits to trim the AWC variation among channels but other approaches may improve on this.

\begin{figure}[!htp]
\includegraphics[width=.95\textwidth]{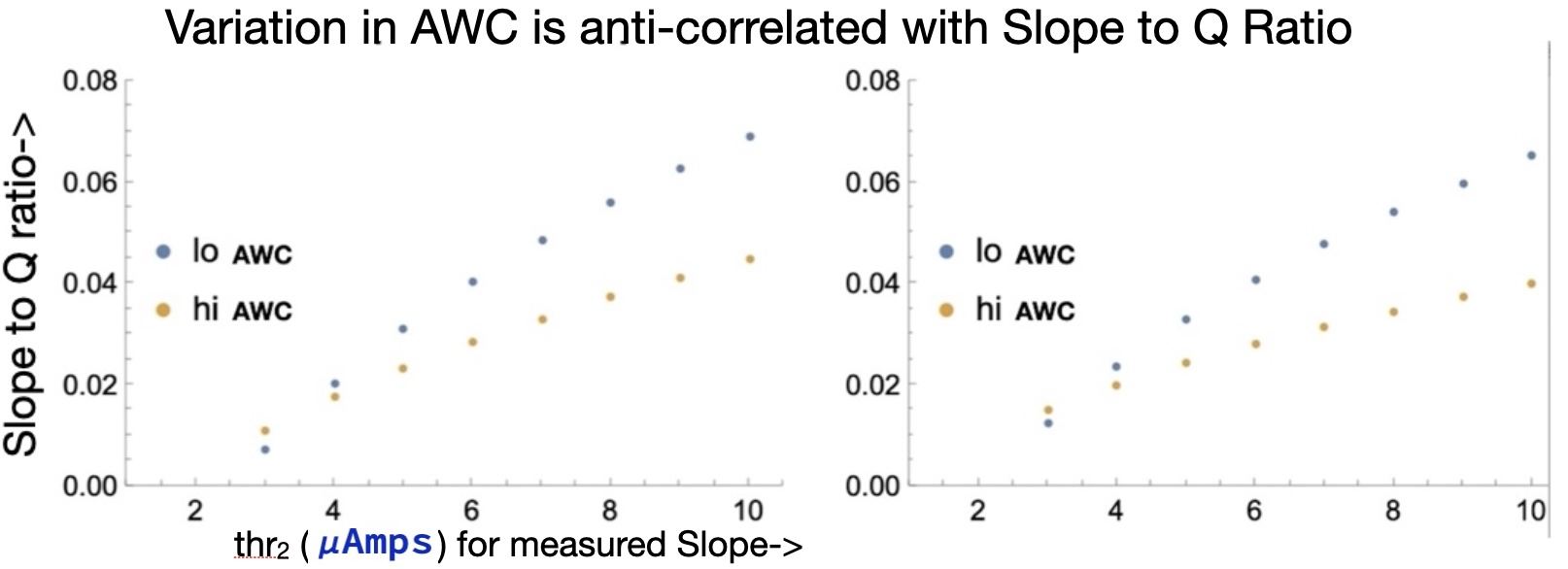}
\caption{We find that although the AWC obtained from fits as in Figure 5 do vary from channel to channel  (``lo AWC" vs ``hi AWC" in the distribution of Figure 6 (Left)), the calibration data contain information to reduce this variation by a factor of 2. 
The corresponding correction factor reducing small channel-to-channel differences is given in eqn. 4.1.}
\label{fig:vary}       
\end{figure}

	This observation leads to a more refined prediction of individual channel AWC to be applied as in eqn. 2.4, where we refer to the linear fit in Figure 6(right). Since the rms spread about
the linear fit varies by only rms$\sim 13\%$ we now have an acceptable Day-1 calibration once the individual channel slope vs. Q tables are accumulated in situ. 

	Then, for our particular case of the TOFHIR, we use the more precise determination of the walk coefficient obtained from the calibration data set and the fit of Figure 6 (right):

\begin{equation}
AWC_{i}=< rawAWC> - 15.8(\frac{1.}{b_{i}}-\frac{1.}{<b>}))
\end{equation}

,where i is the channel index and b$_i$ was obtained from the fit in Eqn. 4.1. The mean value- <rawAWC> -is the average value found in this study ($\sim 30 \mu$Amp).
	
	Having in hand the fixed AWC for each channel, we apply the linear walk correction for events in a given channel:
\begin{equation}
t_{corrected,event}=t_{threshold,event}- AWC\times\frac{1.}{Slope_{event}}
\end{equation}

 ,utilizing the Amplitude (Q) measured in the event and the mapping to Slope from the calibration lookup table.

\section{Discussion}

	For signal amplitudes corresponding to the useful range for BTL response to charged particles the walk correction developed in this study reduces to determining a single 
coefficient for each channel. This linear dependence on the (inverse of) the pulse slope certainly simplifies the calibration in itself.

	In a system, such as the one we are discussing, where the slope correspondence is available for each event, and with a clean environment such as that of ALICE, one could certainly use the ALICE (traditional) approach to determine the corresponding coefficient.
	
	However the HL-LHC environment is certainly more complex than the one experienced by ALICE in Runs 1 and 2 of the LHC.
So in this study we go one step further to achieve a stand alone calibration which doesn't require knowledge of the rest of the events producing hits in the timing array.

	We have seen that the critical feature of the CMS BTL readout is the ability to capture a correspondence between the Signal charge of a hit with a corresponding pulse 
slope at threshold. In the BTL case the slope determination from T1 and T2 threshold times is of limited precision (see Appendix) but using a calibration run that generates hits with
of order 1k or so samples of slope per amplitude bin a sufficiently precise look up table is generated.

	This calibration data set would be acquired periodically to adjust for changes in SiPM gain, for example. The rest of the time the second threshold measurement (thr$_2$) serves no
role and can be dropped. The calibration data serves 2 purposes:

\begin{enumerate}
\item As was shown in Figure 8, different channels of the TOFHIR ASIC differ in the slope evolution with Q. In our study we used the calibration lookup tables (relating slope to Q) to determine a channel dependent AWC coefficient:
\begin{equation}
Slope_i=a_i+b_i\times Q
\end{equation}
and then use the fitted value of b to trim the walk coefficient:
\begin{equation}
AWC_i=<AWC>-\kappa\times(\frac{1}{b_i}-\frac{1}{<b>})
\end{equation}
	,where $\kappa=15.6$ is found from the data in Figure 8. This trimming procedure was useful in our case as it corrected the $22\%$ spread in the AWC coefficients derived from 
the procedure illustrated in Figure 5 to $13\%$.

\item
The direct application of the calibration lookup tables is, of course, to generate the event-by-event walk correction for each channel using eqn. 4.3.

\end{enumerate}

	In conclusion, we have shown that for two example systems (the linear one and the BTL case) the capability to capture slope at threshold leads to a viable stand alone procedure
for walk calibration. First of all we reduce the calibration to determining a single parameter. In the linear case the rest is straightforward. In the more complex case of BTL, The single 
parameter appears to have a small spread ($22\%$ rms)- perhaps useful as a starting point for Day-1 setting. More importantly, for the BTL case, the key to correcting for this small
spread is captured in the calibration data. We find for our data set that this reduces the effective spread in correction parameters to $13\%$.

	The procedure we have outlined certainly enables an initial setting for walk correction without the complexity of the standard procedure. The obvious question that arises is whether or not this is good enough. From the present study, it would seem that further correction could be useful once all of the tools are available. For example, it may turn out that the
trimming procedure for the AWC coefficients is really only good to roughly $10\%$. In that case the overall walk correction for the highest pulse heights (the Landau tail of the distribution)
could be off by of order 2$\times10$ picoseconds or more. A larger data set will be useful to see if the trim procedure can be done more accurately.

	Only with a more comprehensive data set can we say for sure to what extent this calibration procedure will need to be followed up with full event reconstruction. What one would like to see, of course, is that all channels achieve a correction as perfect as the case of Figure 9. For this channel, as in most others, the residual walk is insignificant compared to other terms in the time resolution.

	The next step is clearly to study a larger data set in a particle beam, where we will include several ASICS and replicate the spread in angles and positions of tracks. This effort is currently under way in CMS. Based on the larger data set, the extent to which this procedure satisfies the calibration requirements for the experiment will be evaluated. What we can conclude today is that it will go most of the way to removing amplitude walk and may eliminate the need for a more complex task.


	
	



\begin{figure}[!htp]
\centering
\includegraphics[width=.95\textwidth]{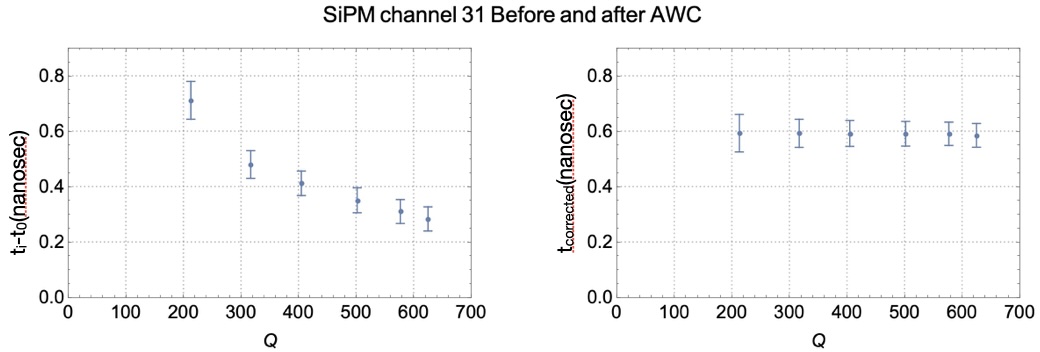}
\caption{ The procedure described in this paper removes the walk observed in the time at threshold (left) illustrated by applying eqn 4.2 to a typical channel (right). }
\label{fig:afterawc}       
\end{figure}

\section{Conclusion}
\label{sec:Conclusion}


	The aim of this exercise has been to show that the initial crucial walk calibration for large timing systems could be obtained with a very restrictive data set-
i.e. an unbiased sample containing only information internal to the system. More sophisticated data including information from the full event- providing a
t$_{0}$ reference - would nevertheless be useful to recover possible channel-to-channel offsets due to propagation delays, etc. But this will be greatly simplified
having removed walk from the data.

\section{Acknowledgement}

	We wish to acknowledge our LIP colleague- M.~Gallinaro for a careful reading of the manuscript and the folks at PETsys, SA for making their test stand available to us. This work was partially supported, through Fermilab, by the US-CMS upgrade program.

\section{Appendix: Future CMS-specific Development}

	Preparation for the commissioning of the CMS BTL includes testbeam exposures of several modules and with different operating conditions- such as SiPM gain.
With the benefit of this large data set it should be possible to determine the applicability of this calibration procedure throughout the lifetime of the experiment- not just for Day-1 conditions. The operating bias and gain of the SiPMs will evolve during the experiment due to SiPM radiation damage and the consequent increase of Dark Count noise.
On the other hand, the TOFHIR ASIC will be maintained at constant temperature and has not exhibited changes in performance up to the anticipated radiation dose. 

	Perhaps more importantly, such a large data set may lead to improved strategies for the AWC coefficient trimming. The simple linear fit of eqn. 5.1 to derive the trim is only a first stab at capturing the subtle differences in ASIC channel behavior. A simple estimate for the target AWC trim would be to achieve a spread of $\leq5\%$. If this is not achievable the
calibration procedure proposed here would, nevertheless, be a useful starting point for a more complex final calibration.

\section{Appendix: Use of Slope Measurement to Mitigate Effects of Dark Count Noise on Time Jitter}

	In an earlier note \cite{dcr} we discussed the possible utility of the slope measurement for reducing time jitter due to dark count noise in SiPMs. This would require event-by-event measurement of the slope. The super-posed 1/f noise (characteristic of dark count noise) \cite{MOS} would disrupt the correlation between slope and amplitude that is highlighted
in the current study. The measured error in slope (relative to expected) was used to mitigate time jitter due to dark noise.  

	This could be a useful tool in future applications of the slope measurement. However, in the case of the TOFHIR, the input electronic bandwidth is such that the
event-by-event slope resolution is degraded by high frequency front-end noise to $\sim 13\%$ as discussed above. In the earlier note \cite{dcr} full waveforms were captured and analyzed with a hard software upper limit on bandwidth at $\sim 300 $ MHz,  which indicates a possible path for future exploitation of this tool.

\end{document}